\documentclass[prl,twocolumn,aps,showpacs,nofootinbib,floatfix,superscriptaddress]{revtex4-1}
\usepackage{amsmath}
\usepackage{amssymb}
\usepackage{graphicx}
\usepackage{verbatim}
\usepackage[colorlinks=true,linkcolor=blue,citecolor=blue,urlcolor=blue]{hyperref}
\usepackage{txfonts}
\usepackage{xr}
\usepackage{hyperref}

\begin{document}

\title{Enhancement of two-photon transitions with tailored quantum light}
\author{Frank Schlawin}
\email{frank.schlawin@physik.uni-freiburg.de}
\affiliation{Physikalisches Institut, Albert-Ludwigs-Universit\"at Freiburg, Hermann-Herder-Stra\ss e 3, 79104 Freiburg, Germany}
\author{Andreas Buchleitner}
\affiliation{Physikalisches Institut, Albert-Ludwigs-Universit\"at Freiburg, Hermann-Herder-Stra\ss e 3, 79104 Freiburg, Germany}
\affiliation{Freiburg Institute for Advanced Studies, Albert-Ludwigs-Universit\"at Freiburg, Albertstra\ss e 19, 79104 Freiburg, Germany}
\date{\today}

\begin{abstract}

We examine a fundamental problem in quantum optics: What is the optimal pulse form to drive a two-photon-transition? 
We show that entangled photons in general do so more efficiently than optimal classical pulses, and
provide the first example of enhanced coherent control through quantum correlations. In ensembles of collectively driven 
two-level systems, such enhancement requires non-vanishing interactions. 
\end{abstract}

\maketitle

The nonlinear interaction between faint light and matter on a single atom/molecule and few-photon level is of great fundamental and practical interest: While Gedanken experiments involving single photons and single quantum emitters have recently come within reach of experimental verification \cite{Sandoghdar1, Sandoghdar2, Sandoghdar3}, similar applications in single molecule spectroscopy may unravel the quantum dynamics of photo-sensitive materials beyond the ensemble average \cite{vanHulst}, and promise to elucidate, for instance, the role of fluctuations in energy transport \cite{Walschaers, Spiros}. Similarly, few-photon experiments in nonlinear media are able to create an effective interaction between photon pairs, and are employed to construct all-optical transistors \cite{Lukin1,Lukin2, Dayan1, Lukin3, Rempe1, Rempe2}. Clearly, the feeble probe and signal fields in such experiments pose a formidable challenge: In order to detect the typically very weak nonlinear effects at small photon numbers (on the order of one), one seeks to, either, optimize the nonlinearity of the optical medium, or to manipulate the light fields: The former route includes the cavity-enhanced coupling of light to the medium \cite{Haroche1, Walther1, Ebbesen}, or the enhancement of the medium's nonlinearity by additional strong driving fields \cite{Lukin1, Lukin2}, large dipoles in highly excited Rydberg states \cite{Xiao1, Hofferberth}, or molecular design of target molecules \cite{Lanco1, Champagne}. In the latter route, strong focussing of the light beams, $i.e.$ the choice of a suitable geometry, was exploited to detect the nonlinear response on a single molecule level \cite{Sandoghdar1, Sandoghdar2, Sandoghdar3}.

Here we present an alternative approach to enhance nonlinear interactions specifically in the highly relevant regime of weak intensities, by employing ideas from coherent control \cite{Gerber1, Silberberg2, Brumer}. Whereas coherent control originally employs strong classical pulses to manipulate the interference between different excitation pathways, recent advances in the manipulation of the two-photon wavefunction of time-frequency entangled photons \cite{Lukens13, Bernhard13, Bessire14, Gigan15} open the possibility to apply coherent control techniques to interactions on the single-photon level. 
We will demonstrate that quantum-mechanical entanglement between photons can enhance the efficiency of the photons in carrying out a given task beyond the classically achievable limit. Incidentally, our results challenge commonly held beliefs in coherent control, according to which the quantum nature of light is detrimental \cite{Brumer}, and open a new avenue for exploiting quantum correlations in the control of molecular processes.

Specifically, we examine a prototypical example of nonlinear light-matter interaction - the (resonant) two-photon absorption in a three-level system driven by either photon pairs \cite{Silberberg1} or famished ``classical" coherent light pulses \cite{Silberberg3}. Using a variational approach, we derive optimal pulse forms for classical and quantum two-photon states, in order to excite the target state $f$ at a given time $t$. Our analytical results demonstrate that the optimal pulse form is, in general, given by an entangled two-photon state. Only if the matter system consists of an ensemble of noninteracting two-level atoms, classical light can perform as well as quantum light. 

\begin{figure}[t]
\centering
\includegraphics[width=.49\textwidth]{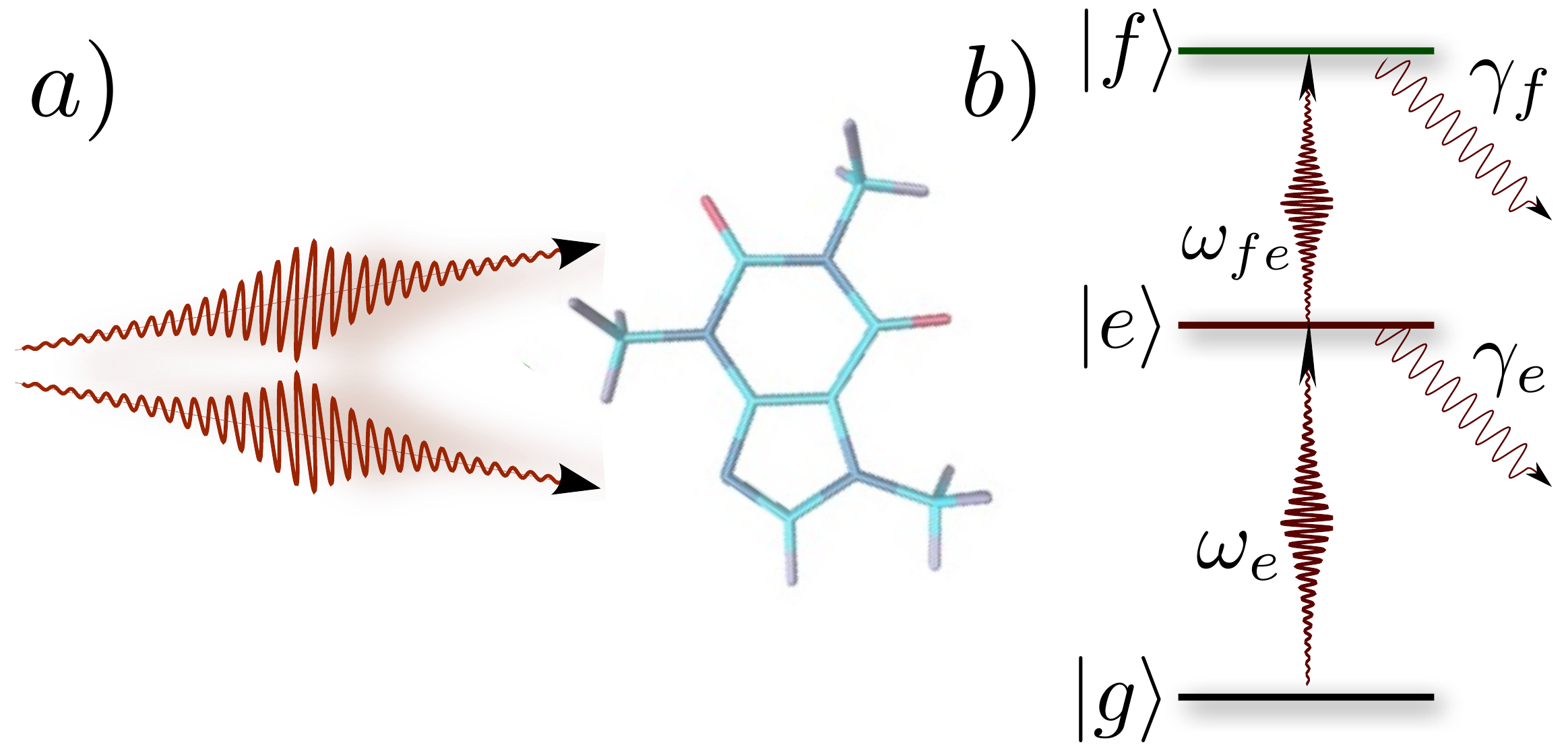}
\caption{(color online) a) A pair of light beams interacts with a two-photon absorbing medium, represented here by a caffeine molecule. b) The considered level scheme with transition frequencies $\omega_e$ and $\omega_{fe}$, as well as excited state lifetimes $\gamma_e$ and $\gamma_f$.}
\label{fig.levelscheme}
\end{figure}
\textit{Two-photon absorption.---}We consider a system of three electronic states with a ground state $g$, intermediate states $e$, and target state $f$ as depicted in Fig.~\ref{fig.levelscheme}b). The system is illuminated by two (possibly quantum) light fields, described by field operators $\hat{E}_{1 , 2} (\omega)$. It is initialized in the ground state, and our goal consists in maximizing the population in state $f$ at time $t$ by controlling the joint quantum state $\psi$ of the two light fields. 
To this end, we consider the functional
\begin{align}
J [\psi] &= \langle \hat{p}_{f} (t)  \rangle_{\psi}- \lambda \left( \big\langle \hat{n}_1 \hat{n}_2 \big\rangle_\psi - N^2 \right). \label{eq.functional-T_fg}
\end{align}
The first term in Eq.~(\ref{eq.functional-T_fg}), $\langle \hat{p}_f (t) \rangle_\psi$, denotes the propability of the system to be in the target state $f$ at time $t$. The expectation value is taken with respect to the fields' quantum state $\psi$, which drives the two-photon transition. The second term in Eq.~(\ref{eq.functional-T_fg}) introduces a Lagrangian multiplier $\lambda$ to enforce the constraint that the product of the mean photon numbers be $N^2$ photons, with the photon number operator given by $\hat{n}_j = \int \! d\omega \; \hat{E}^{\dagger}_j (\omega) \hat{E}_j (\omega)$. 

Since we focus on the interaction with weak fields, we obtain $p_f (t)$ perturbatively, $\langle \hat{p}_f (t) \rangle_\psi = \langle \hat{T}^{\dagger}_{fg} (t) \hat{T}_{fg} (t) \rangle$, where, at second order in the perturbation, the transition amplitude between $g$ and $f$ reads \cite{Cohen-Tannoudji1}
\begin{align}
\hat{T}_{fg} (t) &= \int d\omega_a \int d\omega_b \; T_t (\omega_a, \omega_b) \hat{E}_2 (\omega_b) \hat{E}_1 (\omega_a), \label{eq.two-photon-amplitude}
\end{align}
with the function
\begin{align}
T_t (\omega_a, \omega_b) &= \left(\frac{E_0}{\hbar} \right)^2 \sum_e \left(  \frac{\mu_{ge}}{\omega_a - \omega_e + i \gamma_e} + \frac{\mu_{ge}}{\omega_b - \omega_e + i \gamma_e} \right) \notag \\
&\times \frac{\mu_{ef}}{\omega_a + \omega_b - \omega_f + i \gamma_f}  e^{- i (\omega_a + \omega_b ) t} \label{eq.matter-response}
\end{align}
describing the matter response to the absorption of photons with frequencies $\omega_a$ and $\omega_b$. $\mu_{ge}$ and $\mu_{ef}$ are the dipole moments connecting adjacent electronic states, $\gamma_e$ and $\gamma_f$ are the inverse lifetimes of the respective electronic levels, and $E_0$ the field normalization \cite{MandelWolf}. The summation over $e$ includes several possible intermediate states. Our subsequent analysis is restricted to the simplest case of only one single state, but all the results may be straightforwardly generalized to a manifold of $e$-states. Each excitation of the matter is accompanied by a photon annihilation at the corresponding frequency, rendering Eq.~(\ref{eq.two-photon-amplitude}) a field operator acting on the joint space of the two fields $1$ and $2$.

We further note that, in writing Eq.~(\ref{eq.matter-response}), we allow for pulses extending infinitely long in time, which simplifies our subsequent discussions. The (more involved) general response for finite interaction times will be given elsewhere.
Eq.~(\ref{eq.matter-response}) is symmetric with respect to the exchange $\omega_a \leftrightarrow \omega_b$, which means that we assume each field couples to both transitions. In a situation where, for instance, field $1$ only drives the $g - e$ transition, and field $2$ the $e - f$ transition, we would have to eliminate the second summand in Eq.~(\ref{eq.matter-response}) to obtain the correct result.

To find the solution of our control problem, we require the first variation of the functional~(\ref{eq.functional-T_fg}) to vanish, $\delta J = 0$, and obtain Fredholm integral equations [see the supplementary material] which define the optimal pulse forms.
These may be solved analytically when changing to an appropriate basis of orthogonal functions: As we shall see, this basis is found in the Schmidt decomposition of the matter response function~\cite{Eberly}
\begin{align}
T_t (\omega_a, \omega_b) &= \sum_k r_k \psi^{\ast}_k (\omega_a) \phi^{\ast}_k (\omega_b), \label{eq.Schmidt-decomposition}
\end{align}
where $\{ \psi_k \}$ and $\{ \phi_k \}$ are orthonormal basis sets, and the $r_k$ are real weights, which may be chosen positive.

\textit{Classical vs. quantum light.---}We first consider the excitation by faint classical light fields with amplitudes $A_1 (\omega_a)$ and $A_2 (\omega_b)$. Since the field correlation function in $\langle p_f (t) \rangle_\psi$ is normally ordered \cite{Mollow}, we may replace the field operators in Eq.~(\ref{eq.two-photon-amplitude}) by the field amplitudes \cite{MandelWolf} 
\begin{align}
\hat{E}_1 (\omega_a) \rightarrow A_1 (\omega_a), \quad \text{and} \quad \hat{E}_2 (\omega_b) \rightarrow A_2 (\omega_b). \label{eq.phi_classical}
\end{align}
The optimal frequency distributions $A_1$ and $A_2$, Eqs.~(S5) and (S6), are given by the eigenfunctions pertaining to the largest singular value $r_1$ in Eq.~(\ref{eq.Schmidt-decomposition}), $i.e.$  $A_1 = \sqrt{N} \psi_1$ and $A_2 = \sqrt{N} \phi_1$. Using the orthonormality of the Schmidt basis $\{ \psi_k \}$ and $\{ \phi_k \}$, we may readily carry out the frequency integrations in Eq.~(\ref{eq.two-photon-amplitude}), and the maximal classical probability to excite the target state $f$ at time $t$ then reads 
\begin{align}
p_{f, \text{classical}} (t) &= N^2  r_1^2. \label{eq.max-T_fg_classical} 
\end{align}
While this maximal probability is time-independent, the ideal solutions $A_1$ and $A_2$ depend on the final time $t$ through the time dependence of $T_t$ in Eq.~(\ref{eq.Schmidt-decomposition}).

The simple factorization of the operator-valued argument $\hat{E}_2 (\omega_b) \hat{E}_1 (\omega_a)$ of Eq.~(\ref{eq.two-photon-amplitude}), as implied by Eq.~(\ref{eq.phi_classical}), is no longer possible when describing general quantum states of the field. In particular, when dealing with a general two-photon state, we need to replace $\hat{E}_2 (\omega_b) \hat{E}_1 (\omega_a)$ by the so-called  ``two-photon wavefunction" \cite{Eberly, Gould1}
\begin{align}
\phi_{\text{quantum}} (\omega_a, \omega_b) &= \langle 0 \vert \hat{E}_2 (\omega_b) \hat{E}_1 (\omega_a) \vert \psi \rangle, \label{eq.phi_quantum}
\end{align}
which describes the transition amplitude from the initial state $\vert \psi \rangle$ to the vacuum $\vert 0 \rangle$, upon absorption of photons at frequencies $\omega_a$ and $\omega_b$, respectively. The variation of $\phi_{\text{quantum}}$ yields the optimal two-photon state
\begin{align}
\vert \psi \rangle  &=  \frac{1}{\sqrt{\mathcal{N}}} \int \!\! d\omega_a \int \!\! d\omega_b \;  T^{\ast}_t (\omega_a, \omega_b) \;\vert \omega_a , \omega_b \rangle, \label{eq.Phi_max}
\end{align}
where $\mathcal{N}$ is the normalization constant. The state may be visualized as the time-reversed field emitted by the system in a two-photon decay $f \rightarrow g$, similar to the two-level atoms studied in Refs.~\cite{Stobiska09, Wang11, Kurtsiefer13}.
The ideal excitation probability induced by Eq.~(\ref{eq.Phi_max}) reads
\begin{align}
p_{f, \text{quantum}} (t) &= \sum_{k=1}^{\infty} r_k^2. \label{eq.max-T_fg_quantum}
\end{align}
Dividing the classical result by the mean photon numbers~$N^2$, Eqs.~(\ref{eq.max-T_fg_classical}) and (\ref{eq.max-T_fg_quantum}) demonstrate that, in principle, entangled photons may be tuned such that they drive resonant two-photon absorption processes more effectively than classical laser pulses. 
Due to the joint wavefunction, the phase of each pair of frequency components $(\omega_a, \omega_b)$ in Eq.~(\ref{eq.Phi_max}) is tuned such that the corresponding transition amplitudes - given by Eq.~(\ref{eq.matter-response}) - sum up constructively. With classical light~(\ref{eq.phi_classical}), the phases of $\omega_a$ and $\omega_b$ are determined independently, thus limiting the transition probability.

As a specific example, let us consider excitation along the states $5S_{1/2} \rightarrow 5P_{3/2} \rightarrow 5D_{5/2}$ in Rubidium, with $\omega_e = 780.$~nm, $\gamma_e^{-1} = 26$~ns, $\omega_f - \omega_e = 776$~nm, and $\gamma_f^{-1} = 232$~ns \cite{Safronova}. This particular level scheme was already used in \cite{Silberberg3} to study the influence of the spectral envelope of laser pulses on the two-photon absorption, and for a similar three-state ladder in Rubidium in \cite{Silberberg1} to investigate the absorption of entangled photon pairs. 
We obtain the rather remarkable ratio $p_{f, \text{quantum}} (t) / p_{f, \text{classical}} (t) \gtrsim 8.9$.

These findings naturally lead to the question {\em under which circumstances} quantum light can be more efficient than classical light. To this end, we investigate the enhancement factor
\begin{align}
E (\Gamma) &\equiv \frac{p_{f, \text{quantum}} (t)}{p_{f, \text{classical}} (t)} = 1 + \frac{1}{r_1^2} \sum_{k = 2}^{\infty} r_k^2, \label{eq.enhancement-factor}
\end{align}
where $\Gamma$ denotes the set of parameters which enter in Eq.~(\ref{eq.matter-response}).
\begin{figure}[t]
\centering
\includegraphics[width=.38\textwidth]{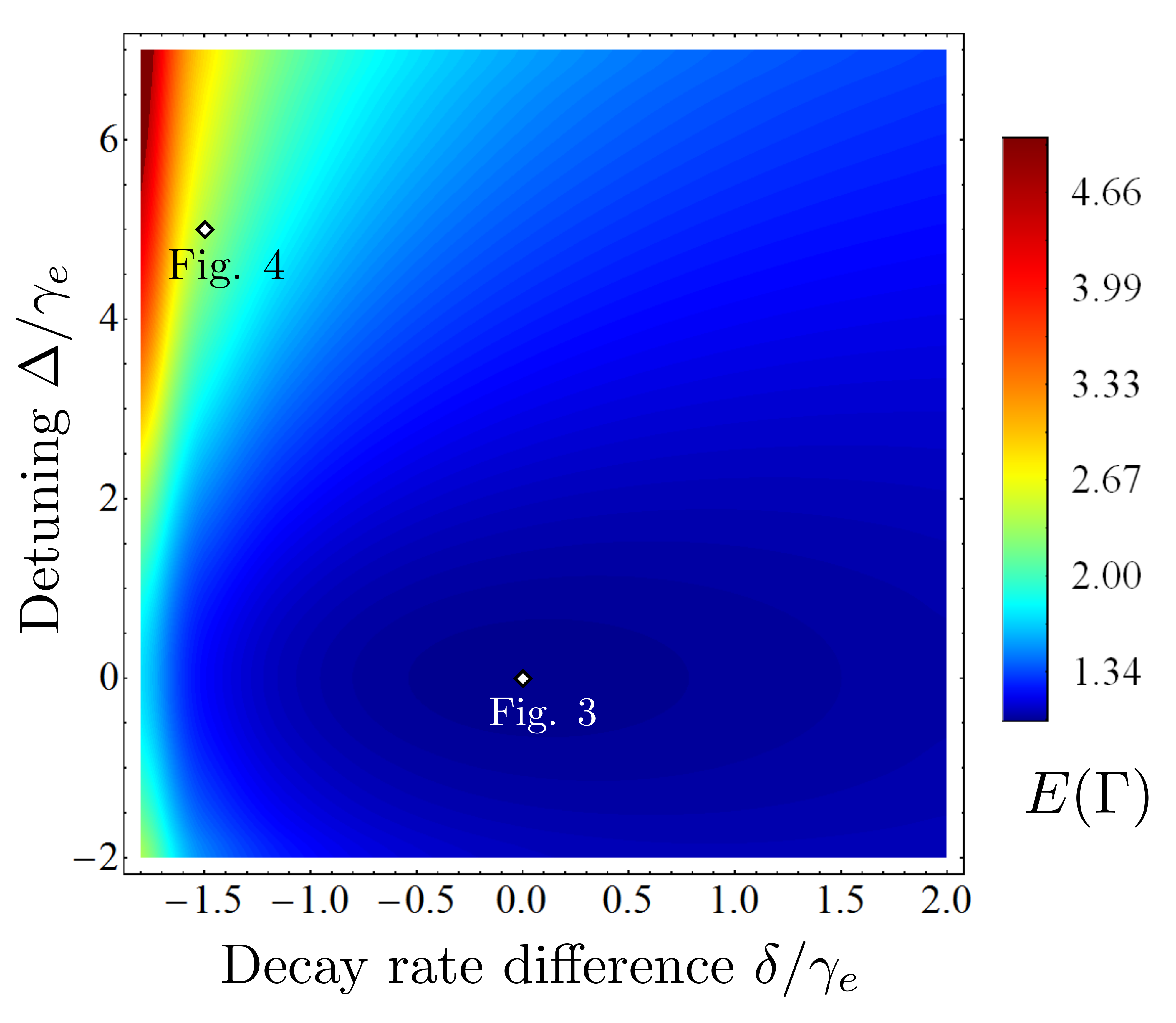}
\caption{(color online) Enhancement factor~(\ref{eq.enhancement-factor}) as a function of the detuning $\Delta = \omega_{fe} - \omega_e$ and of the decay rate difference $\delta = \gamma_f - 2 \gamma_e$, given in units of the inverse lifetime $\gamma_e$ of the intermediate state (see Fig.~\ref{fig.levelscheme}). The result is independent of the dipole moments $\mu_{ge}$ and $\mu_{ef}$, since both can be factored out in Eq.~(\ref{eq.matter-response}).
The rectangles indicate the parameters used in Figs.~\ref{fig.harmonic-pulses} and \ref{fig.anharmonic-pulses}.}
\label{fig.2d-enhancement}
\end{figure}
We evaluate $E (\Gamma)$ as a function of the detuning $\Delta = \omega_f - 2 \omega_e$ of the $e - f$ transition from the $g - e$ transition, and of the deviation $\delta = \gamma_f - 2 \gamma_e$ of the inverse lifetime of the target state from twice the intermediate state's lifetime.
The reason behind this peculiar choice of coordinates becomes evident in Fig.~\ref{fig.2d-enhancement}: The enhancement factor $E (\Gamma)$ forms a convex function of its arguments $\Delta$ and $\delta$, which takes its unique minimum value $E = 1$ ($i.e.$ no enhancement compared to excitation by classical pulses) at $\Delta = \delta = 0$.

To understand this minimum, we note that we can rewrite the two-photon response function for $\Delta = \delta = 0$ as a product of Lorentzian response functions, 
 \begin{align}
&\left[ \frac{1}{\omega_a - \omega_e + i \gamma_e} + \frac{1}{\omega_b - \omega_{e} + i \gamma_{e}} \right] \frac{1}{\omega_a + \omega_b -2 \omega_e  + i 2 \gamma_e} \notag \\
&=\frac{1}{(\omega_a - \omega_e + i \gamma_e) (\omega_b - \omega_{e} + i \gamma_{e})}. \label{eq.factorization}
\end{align}
These Lorentzian factors can be interpreted as single-photon response functions for the transition $g \rightarrow e$, for which the transition amplitude in first-order perturbation theory reads \cite{Cohen-Tannoudji1}
\begin{align}
T_{eg} (t) &= \frac{E_0}{\hbar} \int d\omega \frac{\mu_{eg}}{\omega - \omega_e + i \gamma_e} e^{- i \omega t} A (\omega).
\end{align}
Here, the amplitude $A (\omega)$ may describe, both, a classical field amplitude or a single-photon spectral envelope.
{\em Hence, at $\Delta = \delta = 0$ the two-photon response function factorizes into the product of single-photon response functions, and no correlations between the first and the second absorption event can induce any quantum advantage.} Such a situation occurs, for instance, when the sample system consists of noninteracting two-level atoms, for which entangled two-photon absorption was investigated in Refs.~\cite{Scully2, Richter11, Zhang13}. Our present analysis shows that entanglement cannot enhance the efficiency of the two-photon absorption process in such systems beyond the classical limit: If we consider two two-level atoms with transition frequencies $\omega_e$ and $\omega_{e'}$, and each photon interacts with only one of the two-level atoms, we can modify Eq.~(\ref{eq.factorization}) to write
\begin{align}
&\frac{1}{(\omega_a - \omega_e + i \gamma_e) (\omega_b - \omega_{e'} + i \gamma_{e'})}\notag \\
=&\left[ \frac{1}{\omega_a - \omega_e + i \gamma_e} + \frac{1}{\omega_b - \omega_{e'} + i \gamma_{e'}} \right] \notag \\
&\times \frac{1}{\omega_a + \omega_b - \omega_e - \omega_{e'}  + i ( \gamma_e+ \gamma_{e'})}.
\end{align}
The two-photon response again factorizes, and there can be no enhancement beyond the classical limit due to photonic entanglement. By comparing optimal pulse forms rather than specific models of light in~\cite{Scully2, Richter11, Zhang13}, our analysis unambiguously settles this debate.

\begin{figure}
\centering
\includegraphics[width=.49\textwidth]{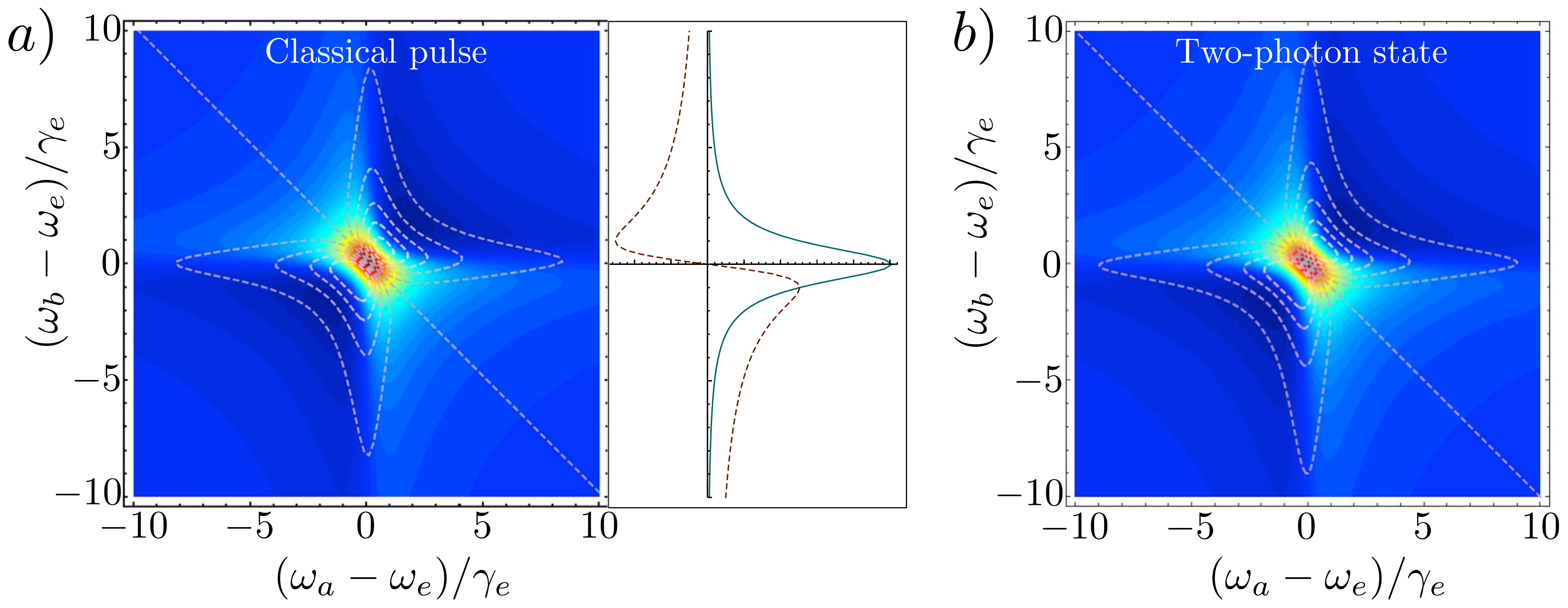}
\caption{(color online) a) Left: Real part of the product $A_2 (\omega_b) A_1 (\omega_a)$ for a three-level system with $\omega_f = 2 \omega_e$ and $\gamma_f = 2 \gamma_e$. The dotted lines indicate equipotential lines of the imaginary part. Right: Real (solid) and imaginary part (dashed) of the field amplitude $A_i (\omega)$. b) Real part of the optimal two-photon state~(\ref{eq.phi_quantum}) for the same matter system as panel a). 
This is the parameter regime with enhancement factor $E = 1$ (see Fig.~\ref{fig.2d-enhancement}).}
\label{fig.harmonic-pulses}
\end{figure}

Since the above discussion also establishes that entangled photons indeed can enhance the two-photon transition probability when either $\Delta \neq 0$ or $\delta \neq 0$, we next investigate the optimal pulse forms in detail. Fig.~\ref{fig.harmonic-pulses} depicts the optimal classical field amplitudes $A_2 (\omega_b) A_1 (\omega_a)$ [panel a)] and two-photon wavefunction $\phi (\omega_a, \omega_b)$ [panel b)] at the minimum in Fig.~\ref{fig.2d-enhancement}, $i.e.$ when the two-photon transition amplitude factorizes as in Eq.~(\ref{eq.factorization}). Clearly, the two pulse forms are identical, as was clear from our theoretical analysis. Since in our present study Eq.~(\ref{eq.matter-response}) was chosen symmetric with respect to the exchange $\omega_a \leftrightarrow \omega_b$, so are the two field amplitudes, $i.e.$ $A_1 = A_2$, and it is sufficient to show one of them in the right inset of panel a). It is simply given by a Lorentzian lineshape $\sim 1 / (\omega - \omega_e - i \gamma_{e})$, $i.e.$ in the time domain by a ``rising exponential" shape. This behavior was to be expected from previous results on single-photon excitation of a two-level atom \cite{Stobiska09, Wang11, Kurtsiefer13}, where a ``rising exponential" temporal shape was shown to couple most efficiently to the matter system. In the case of noninteracting two-level atoms, it is quite intuitive that this pulse shape is still optimal for two-photon excitations, when the two absorption events are not correlated.

\begin{figure}
\centering
\includegraphics[width=.49\textwidth]{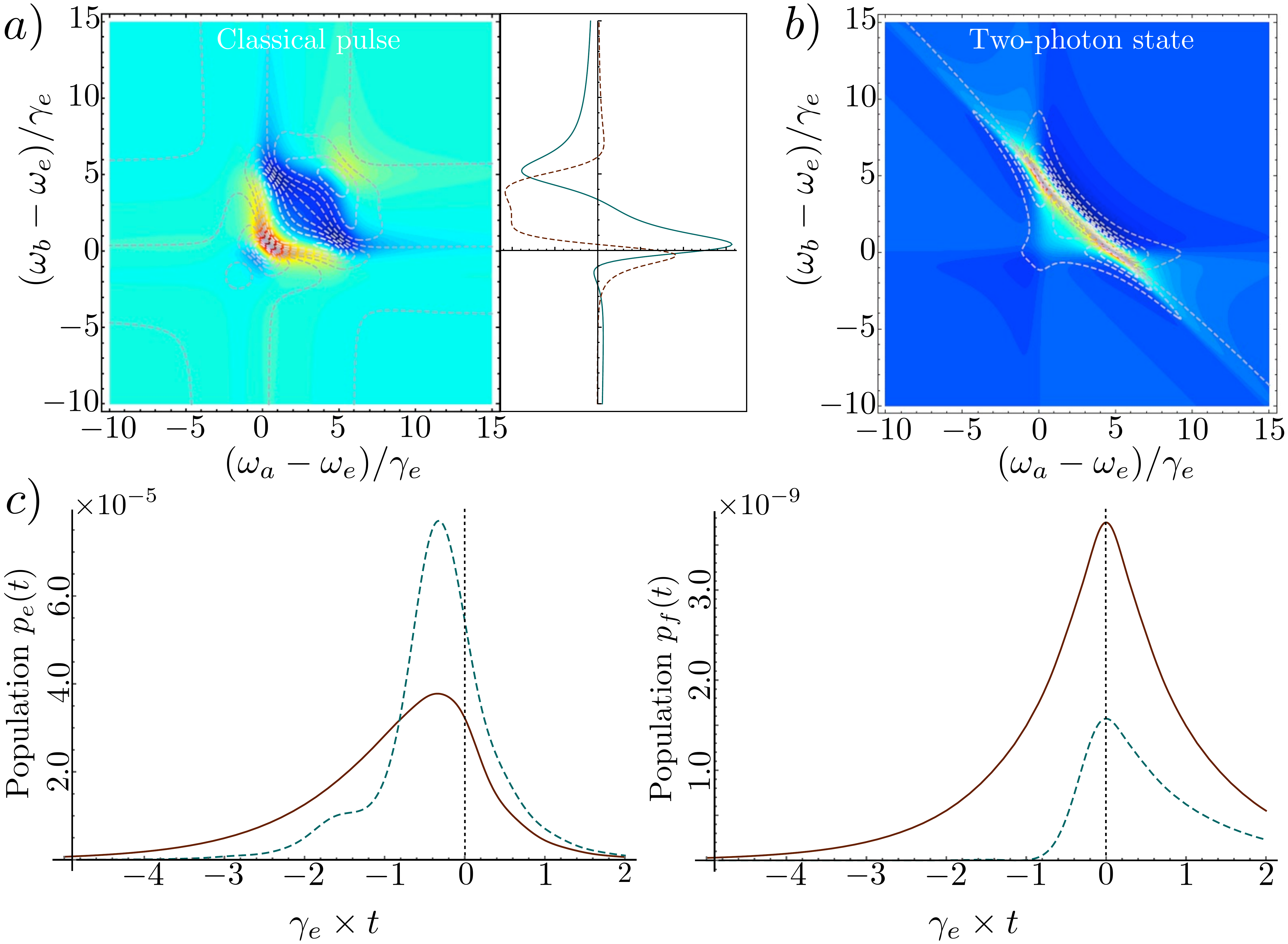}
\caption{(color online) a) Left: Real part of the product $A_2 (\omega_b) A_1 (\omega_a)$ for a three-level system with $\omega_f = 2 \omega_e$, $\gamma_f = \gamma_e / 2$, and $\Delta = 5 \gamma_e$. Right: Real (solid) and imaginary part (dashed) of the field amplitude $A_i (\omega)$. b) Real part of the optimal two-photon state~(\ref{eq.phi_quantum}) for the same matter system as panel a). c) Intermediate state population $p_e (t)$ (left) and target state population $p_f (t)$ (right) created by the interaction with the optimal classical pulse of panel a) (blue, dashed) and the optimal two-photon state of panel b) (red, solid). The dipole moments are chosen $\mu_{ge} E_0 = \mu_{ef} E_0 = 0.01 \hbar / \gamma_e$. The dashed, vertical line indicates $t = 0$, at which point the population in state $f$ is to be maximized.
This parameter regime allows for an enhancement factor $E = 2.4$ (see Fig.~\ref{fig.2d-enhancement}). }
\label{fig.anharmonic-pulses}
\end{figure}

The situation changes substantially when the two-photon state does enhance the transition probability: In Fig.~\ref{fig.anharmonic-pulses}, we depict the case where the second transition energy is larger than the first one, with $\Delta = 5 \gamma_e$, and the target state lives longer than the intermediate state, $\gamma_f = \gamma_e /2$  (as was also the case in our rubidium example above): The optimal classical pulses [panel a)] show a broad structure in the area $(\omega_a, \omega_b) \in (\omega_e \ldots \omega_e + \Delta, \omega_e \ldots \omega_e + \Delta)$. While, at first glance, these optimal pulses appear very different from the harmonic case before, this structure can still be understood by the interference of three Lorentzians: In addition to the harmonic Lorentzian of Fig.~\ref{fig.harmonic-pulses}, one more Lorentzian contributes for, each, the nonvanishing $\Delta$ and $\delta$. 

In contrast, the optimal entangled wavefunction [panel b)] forms a narrow structure along the anti-diagonal. Its width along the diagonal becomes a lot narrower than in Fig.~\ref{fig.harmonic-pulses}a), reflecting the narrow linewidth of the target state $\gamma_f$, whereas the width along the anti-diagonal remains the same, since the lifetime of the intermediate state is unaltered. Due to the detuning $\Delta$ of the $e- f$ transition, it acquires a double peak structure with maxima at $(\omega_a, \omega_b) = (\omega_e, \omega_e + \Delta)$, and at $(\omega_e + \Delta, \omega_e)$, such that the sum of the two frequencies always matches $\omega_f$. In contrast to the classical pulses in panel a), the two-photon wavefunction shows no intensity at $(\omega_e, \omega_e)$, nor at $(\omega_e + \Delta, \omega_e + \Delta)$. This is only possible in the quantum regime, since the absorption of the first photon affects the wavefunction of the other photon, thereby influencing the second absorption event \cite{Schlawin14}. 
The two-photon wavefunction is a clear indicator of strong frequency anticorrelations in the pure two-photon state considered here \cite{Eberly, Kim, Franson}, and it is these correlations that are responsible for the $140$~\% enhancement ($E = 2.4$) of the quantum transition probability. 

To see how the spectral structures of the optimal pulses affect the absorption process, we simulate the time evolution of the excited state populations $p_e (t)$ (left) and $p_f (t)$ (right) in panel c) of Fig.~\ref{fig.anharmonic-pulses} \cite{Oka10}. 
Clearly, while the maximal population in the intermediate state $p_e$ is larger for the case of classical pulses, the maximal target state population $p_f (t = 0)$ upon absorption of entangled pairs exceeds the ``classical" optimal population by a $1.4$-fold - as predicted by the enhancement factor~(\ref{eq.enhancement-factor}) for the current set of parameters.

The broad bandwidth of the classical pulses is reflected in a steep rise of both the intermediate and target state populations near $t = 0$. The intermediate state $e$ decays more rapidly than the target state $f$, such that the optimal classical pulse has to strongly excite this state shortly before the population in $f$ shall be maximized.

In contrast, the populations induced by the absorption of entangled photons rise much slower. Since frequency-anticorrelated photons arrive in pairs \cite{Silberberg2}, the time the system spends in the intermediate state is minimized. It further means that at any point in time there is a finite probability to find the system in the intermediate state. This driving causes a perfectly symmetric progression of the target state population $p_f (t)$ which rises with the same speed it subsequently decays again. This could be anticipated from the optimal state~(\ref{eq.Phi_max}) which is given by the complex conjugate (hence, the time-reverse) of the matter response~(\ref{eq.matter-response}). The inability of the classical pulses to reproduce this symmetric time evolution lies at the heart of the quantum advantage discussed here.

\textit{Summary \& Outlook.---}We investigated optimal pulse forms to efficiently drive a resonant two-photon transition with weak fields. Our study represents - to the best of our knowledge - the first theoretical investigation of coherent control theory in the presence of quantum correlations. Whereas previous studies on coherent control with quantum light focus on the manipulation of interference between excitation pathways \cite{Brumer}, where the quantum properties of light may in fact be detrimental, our study shows that photonic entanglement potentially renders the two-photon absorption process significantly more efficient than classical pulses - except for two-photon absorption in noninteracting systems. Such shaping of the two-photon wavefunction provides a useful application of entangled photons as a spectroscopic probe of photosensitive samples where large photon fluxes are to be avoided \cite{Goodson13, Schlawin13, Raymer13}.
It provides yet another instance in which quantum correlations may enhance a given task compared to classical analogs, and joins the ranks of scenarios such as quantum computation or imaging \cite{quhub}. 

Our introduction of the Schmidt decomposition - a concept from quantum information theory - to analyze the material response function could have applications well beyond the simple systems considered here, as it offers a new perspective on the nonlinear response of (possibly complex) quantum systems. The Schmidt coefficients - very much like in the case of photonic entanglement, where they determine the effective dimension of the Hilbert space available for quantum information purposes \cite{Eberly} - contain information about couplings between resonances in the quantum system. The current approach extracts this information from the optical response of the system, and could thus provide a tool to analyze entanglement with optical measurements, for instance, the interaction-induced entanglement of excitons in complex systems \cite{Daniel}.

\begin{acknowledgments}
\textit{Acknowledgments.} We would like to acknowledge helpful discussions with Stefan Lerch and Andr\'{e} Stefanov, which stimulated the current letter. F.~S. would like to thank the German National Academic Foundation for support.
\end{acknowledgments}

\newpage

\onecolumngrid
\appendix

\renewcommand{\thepage}{S\arabic{page}}
\renewcommand{\thesection}{}
\renewcommand{\thetable}{S\arabic{table}}
\renewcommand{\theequation}{S\arabic{equation}}
\renewcommand{\thefigure}{S\arabic{figure}}
\renewcommand\refname{Supplementary References}
\renewcommand\bibname{Supplementary References}

\section{Integral equations}
\label{sec.integral-equations}

\subsection{Classical fields}
If the matter system is excited by classical fields with envelopes $A_1$ and $A_2$, the field operators are replaced by the classical amplitudes [see Eq.~(\ref{eq.phi_classical})]. Requiring the variation of Eq.~(\ref{eq.functional-T_fg}) to vanish for arbitrary changes $\delta A_1$ and $\delta A_2$ around the optimal pulse forms, we obtain the nonlinear integral equations 
\begin{align}
A_1 (\omega_a) &= \frac{1}{\int d\omega \vert A_2 (\omega) \vert^2} \int d\omega_b \int d\omega'_a \int d\omega'_b \; K (\omega_a, \omega_b; \omega'_a, \omega'_b) A^{\ast}_2 (\omega_b) A_2 (\omega'_b) A_1 (\omega'_a), \label{eq.variation-cl1}\\
A_2 (\omega_b) &=\frac{1}{\int d\omega \vert A_1 (\omega) \vert^2}  \int d\omega_a \int d\omega'_a \int d\omega'_b \; K (\omega_a, \omega_b; \omega'_a, \omega'_b) A^{\ast}_1 (\omega_a) A_2 (\omega'_b) A_1 (\omega'_a), \label{eq.variation-cl2}
\end{align}
with the kernel
\begin{align}
 K (\omega_a, \omega_b; \omega'_a, \omega'_b) &= \frac{1}{\lambda} T_t (\omega'_a, \omega'_b) T^{\ast}_t (\omega_a, \omega_b). \label{eq.kernel}
\end{align}
They are subject to the normalization
\begin{align}
\int \!\! d\omega_a \; \vert A_1 (\omega_a) \vert^2 \int \!\! d\omega_b \; \vert A_2 (\omega_b) \vert^2 &= N^2.
\end{align}
Using the Schmidt decomposition~(\ref{eq.Schmidt-decomposition}) of the matter response function, it becomes apparent that
\begin{align}
A_1^{(k)} (\omega_a) &= \sqrt{N} \psi_k (\omega_a), \label{eq.A_1}\\
A_2^{(k)} (\omega_b) &= \sqrt{N} \phi_k (\omega_b) \label{eq.A_2}
\end{align}
are solutions of the integral equations. Hence, the solutions pertaining to the largest Schmidt eigenvalue $r_1$ maximize the classical two-photon absorption within this set of solutions.

\subsubsection{Maximality of the solutions}
This section proves that the solutions $A_1^{(1)}$ and $A_2^{(1)}$ are in fact the optimal classical pulse forms.

Suppose there are different solutions $\tilde{A}_1 (\omega)$ and $\tilde{A}_2 (\omega)$ of Eqs.~(\ref{eq.variation-cl1}) and (\ref{eq.variation-cl2}). Since the basis sets $\{ \psi_k \}$ and $\{ \phi_k \}$ each form complete sets of orthogonal functions, we may expand $\tilde{A}_1 (\omega) = \sqrt{N} \sum_k c_k \psi_k (\omega)$ and $\tilde{A}_2 (\omega) = \sqrt{N} \sum_k d_k \phi_k (\omega)$. The normalization of the field contribution in Eq.~(\ref{eq.functional-T_fg}) yields (for simplicity, we only consider the case $N = 1$) $\sum_k \vert c_k \vert^2 = \sum_k \vert d_k \vert^2 = 1$, and the two-photon amplitude, upon insertion of the solutions~(\ref{eq.A_1}) and (\ref{eq.A_2}) into the transition amplitude~(\ref{eq.two-photon-amplitude})
\begin{align}
p_f (t) &= \vert N \sum_k r_k c_k d_k \vert^2 \leq \left( N \sum_k  r_k\vert c_k \vert \vert d_k \vert \right)^2 \notag \\
&\leq N^2 r_1^2 \left( \sum_k \vert c_k \vert \vert d_k \vert \right)^2,
\end{align}
where we have used the triangle inequality in the first estimation. To proceed, we invoke the H\"{o}lder-inequality \cite{Abramowitz}
\begin{align}
\sum_k \vert x_k y_k \vert \leq \left( \sum_k \vert x_k \vert^p \right)^{1/p} \left( \sum_k \vert y_k \vert^q \right)^{1/q},
\end{align} 
to write (with $p = q = 2$),
\begin{align}
p_f (t) &\leq N^2 r_1^2 \sum_k \vert c_k \vert^2 \sum_k \vert d_k \vert^2 = N^2 r_1^2,
\end{align}
where we recall $N^2 r_1^2$ is the optimal result~(\ref{eq.max-T_fg_classical}) of the pulses $A_1^{(1)}$ and $A_2^{(1)}$. Hence, $A_1^{(1)}$ and $A_2^{(1)}$ are indeed the optimal solutions. 

Similarly, an incoherent mixture of different eigenfunctions $\psi_k$ and $\phi_k$ with weights $\tilde{p}_k$ cannot enhance the maximal transition probability~(\ref{eq.max-T_fg_classical}): We obtain
\begin{align}
p_f (t) &= N^2 \sum_k \tilde{p}_k r_k^2 \leq N^2 r_1^2 \sum_k \tilde{p}_k = N^2 r_1^2.
\end{align}

\subsection{Photon pairs}
In case the initial state of the light field $\psi$ is given by a two-photon state, the field enters in Eq.~(\ref{eq.two-photon-amplitude}) with its two-photon wavefunction, $\phi (\omega_a, \omega_b) = \langle 0 \vert E (\omega_b) E (\omega_a) \vert \psi \rangle$ [see Eq.~(\ref{eq.phi_quantum})]. In an analogous calculation to the derivation of Eqs.~(\ref{eq.variation-cl1}) and (\ref{eq.variation-cl2}), we require the variation of the functional to vanish for arbitrary changes $\delta \phi$ around the optimal solution. We then obtain the homogeneous Fredholm equation \cite{Abramowitz}
\begin{align}
\phi (\omega_a, \omega_b ) &= \int d\omega'_a \int d\omega'_b \; K (\omega_a, \omega_b; \omega'_a, \omega'_b) \phi (\omega'_a, \omega'_b), \label{eq.variation1}
\end{align}
with the kernel given in Eq.~(\ref{eq.kernel}), which determines the optimal two-photon wavefunction. It is solved by
\begin{align}
\phi_{\text{max}} (\omega_a, \omega_b) &= \frac{1}{\sqrt{\mathcal{N}}} \sum_k r_k \; \psi_k (\omega_a) \phi_k (\omega_b) \label{eq.Phi_max2}
\end{align}
with the (state) normalization
\begin{align}
\mathcal{N} &= \int d\omega_a \int d\omega_b \; \vert T_t (\omega_a, \omega_b) \vert^2 \\
&= \frac{2 \pi^2 \mu_{ge}^2 \mu_{ef}^2}{\hbar^4 \gamma_e \gamma_f}. \label{eq.N}
\end{align}
This corresponds to the two-photon state~(\ref{eq.phi_quantum}) of the main text.

The state normalization also represents the maximal population in the target state $f$. Using Eq.~(\ref{eq.Phi_max2}), the population $p_f (t)$ reads
\begin{align}
p_f (t) = \vert T_{fg} \vert^2 = \frac{1}{\mathcal{N}} \vert T^{\ast}_t (\omega_a, \omega_b) \vert^2 = \mathcal{N}. \label{eq.N2}
\end{align}

\end{document}